\documentstyle{l-aa} 
\font\sixrm=cmr6 
                            
\newcommand\ion[2]{\hbox{#1\,{\sixrm #2}}} 
\newcommand{\teff}{$\rm T_{eff}$} 
 
\newcommand\loghe{${\rm \log{\frac{N_{He}}{N_H}}}$ } 
\newcommand\logg{${\rm \log g}$} 
\newcommand{\Msolar}{\mbox{\,$\rm M_{\odot}$}}        
\newcommand{\pgsie}{PG\,1704+222} 
\newcommand{\pgdre}{PG\,1323$-$086}
\newcommand{\bddre}{BD\,+33$^\circ$2642} 
\begin{document} 
\thesaurus{7(08.01.1, 08.05.1, 08.16.4, 08.05.3)} 
\title{\pgdre\ and \pgsie\ -- two post-AGB stars at high galactic  latitudes} 
\author{S. Moehler  \inst{1}  
\thanks{Visiting Astronomer, German-Spanish Astronomical Center,  Calar Alto,
operated by the Max~Planck-Institut f\"ur Astronomie  jointly with the Spanish
National Commission for Astronomy}  
\and U. Heber \inst{1} }
\offprints{S. Moehler} 
\institute{Dr. Remeis-Sternwarte, Astronomisches Institut der Universit\"at
Erlangen-N\"urnberg, Sternwartstr. 7, 96049 Bamberg,  Federal Republic of
Germany} 
\date{Received , accepted }  
\maketitle

\begin{abstract}  Two high galactic latitude B-type stars, \pgdre\ and \pgsie ,
are analysed from low and medium resolution optical spectra and Str\"omgren 
photometry. A differential abundance analysis for He, C, N, O, Mg, Al, and Si 
reveals  that the He abundance is close to solar
while metal underabundances relative
to the  solar value of typically 1.3 and 1.2 dex for \pgdre\ and \pgsie ,
respectively, are found. For both stars, carbon is even more depleted. The
atmospheric parameters are consistent with  evolutionary tracks for stars
evolving from the asymptotic giant branch  (AGB) with a stellar mass of
0.55~\Msolar. The anomalous compositions are compatible with those  of other
high galactic latitude post-AGB stars. Hence,  \pgdre\ and \pgsie\ are low mass
post-AGB stars in an  evolutionary stage between those of A- and F-type
supergiants of low mass  and central stars of planetary nebulae.  From
kinematic data and  distances a population~II membership is probable for both
stars.

\end{abstract}
\section{Introduction}

In a long-term project we have followed-up spectroscopically on  hot subdwarf
candidates from the ``Palomar-Green Catalogue of Ultraviolet Excess Stellar
Objects'' (Green et al., \cite{gree86}). While most objects turned out to be
hot subluminous stars (Moehler et al., \cite{mohe90b}, Theissen et al.,
\cite{thmo93}) or blue horizontal branch stars (Schmidt et al., \cite{schm92})
a few candidates showed surface gravities too low for such  stars and more
typical for giant stars. In addition these objects exhibit normal helium line
strengths, again in contrast to the usually helium weak-lined hot subdwarfs.

The normality of the H/He spectrum might indicate that these stars are normal,
i.e. massive giants in a strange place, very far from the  galactic disk. In
this case they were either created in the disk and thrown out (cf. Leonard,
\cite{leon91}) or they were born in the halo  (which could be explained by the
model of the ``galactic fountain'',  Dyson \& Hartquist, \cite{dyha83}). On the
other hand, such stars could be post-AGB stars on their way from the 
asymptotic giant branch (AGB) to the white dwarf domain that spectroscopically
mimic massive stars.  Since this part of the evolution proceeds very fast 
(10$^4$ years, Sch\"onberner, \cite{scho83}), such objects should be rather 
rare. Indeed, not many have been found and analysed so far (Conlon et al., 
\cite{codu91}, \cite{conl93}). 

Since a metallicity analysis offers a good possibility to distinguish between
young, massive OB stars and post-AGB stars we decided to get higher resolution
(about 1~\AA) spectra of some of these stars. Here we report the results for
two objects, \pgdre\ and \pgsie . Lower resolution spectra of \pgsie\ have
already been discussed by Conlon et al.  (\cite{conl93}). In addition to these
stars we observed spectra of well known and analysed main-sequence and
sub-giant OB stars to allow direct comparisons between the spectra. In Section
2 we describe the observations and the reduction of the data. Sections 3 and 4
describe the determination of atmospheric  parameters and abundances. Section 5
gives the kinematic data of the stars,  while Section 6 gives a discussion of
the results.

\section{Observations and reduction}

The observations and reduction of medium resolved spectra are described in
Moehler et al. (\cite{mohe90a}, \pgdre ) and Conlon et al.  (\cite{conl93},
\pgsie ). The higher resolved data were obtained with the  Cassegrain Twin
Spectrograph of the 3.5m telescope at the Calar Alto  observatory, using a
slit-width of 2.1$\arcsec$ and a mean dispersion of 18~\AA/mm in the blue and
of 36~\AA/mm in the red channel.  Both channels were equipped with Tektronix
CCDs with 1024 $\times$ 1024 pixels of 24 $\mu$m size. Read-out-noise and gain
were 9~e$^-$ and 4.5~e$^-$/ADU in the blue and 8~e$^-$ and 4.5~e$^-$/ADU in the
red channel. With this setup we covered the wavelength range 3850~\AA\ --
4710~\AA\ in the blue channel with two exposures and in the red channel we used
the wavelength range 5850~\AA\ -- 6720~\AA. The mean FWHM of the calibration
lines was about 0.9~\AA\ and 1.8~\AA\ for the  blue and red channel,
respectively, and represents the typical  spectral resolution of the system.

The observations and data reduction were performed as described by Moehler  et
al. (\cite{mohe95}) except that we did not use any optimal extraction 
algorithm, as the spectra were tilted on the CCD.
Instead we summed up the rows of each spectrum  after sky-subtraction and tried
to get rid of cosmic events by comparing several spectra of the same object.
Finally the spectra were corrected for the Doppler shifts  and coadded. The
resulting mean radial velocities are listed in Table~5. To give an idea of the
quality of our spectra we show part of them in  Fig.~1.  \begin{figure} 
\includegraphics{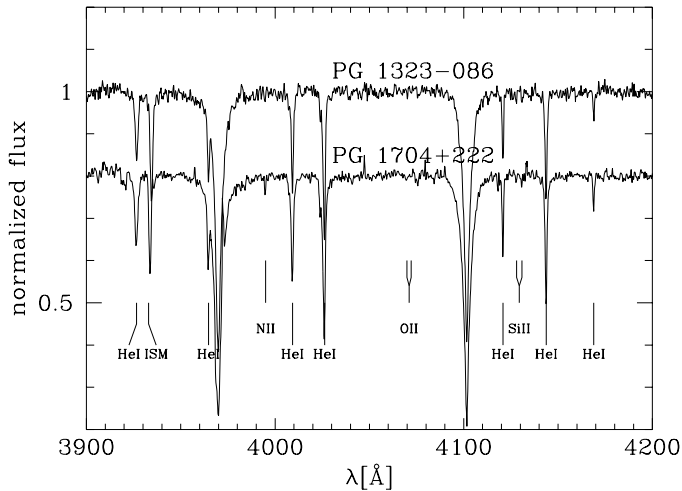} \vspace{6.5cm} \caption[]{Part of the blue spectra of \pgdre\ and
\pgsie . Those spectral lines that could be identified are marked.}
\end{figure}

\section{Physical Parameters}

For both stars Str\"omgren photometry is available from the literature
(Wesemael et al. \cite{wese92}), which can be  used in combination with the
observed Balmer line profiles to derive effective temperatures and surface
gravities for these objects. 

\teff\ and \logg\ were derived from the reddening-free indices [c$_1$]
and [u$-$b].  Since post-AGB stars tend to
show sub-solar  metallicities (see McCaus\-land et al., \cite{mcco92}) we used
theoretical colours  for metallicity $-$1  (Napiwotzki \& Lemke, priv. comm., a
recent extension of the  Napiwotzki et al., \cite{nasc93}, calibration to lower
metallicities). We also fitted theoretical ATLAS9 Balmer line profiles for
metallicity $-$1 and solar helium abundance  to the H$_\delta$ to H$_\beta$
lines (keeping \teff\ fixed), which yielded another set of \teff , \logg\ 
values. The intersection of the curves shown in Fig. 2 resulted in the
parameters listed in Table~1. The \ion{He}{I} lines are fitted very well by
these models, suggesting a  close to solar helium abundance.

As a cross check we also derived the physical parameters from the line 
profiles only, using the routines of Bergeron et al. (\cite{besa92}) and Saffer
et al. (\cite{saff94}), which employ a $\chi^2$ test. We computed model
atmospheres using ATLAS9 (Kurucz 1991, priv. comm.) and  used the LINFOR
program (developed originally by Holweger, Steffen, and Steenbock at Kiel
university)  to compute a grid of theoretical spectra, which include the Balmer
lines H$_\alpha$ to H$_{22}$ and \ion{He}{I} lines. We fitted H$_\epsilon$ to
H$_\beta$ and the \ion{He}{I} lines $\lambda\lambda$ 4026, 4121, 4388, 4437,
4471 simultaneously.  As can be seen from Table~1 the values  are very similar
for both procedures. 

A third approach to derive \teff\ is to use the \ion{Si}{II}/\ion{Si}{III} 
ionization equilibrium, which is applicable only to \pgdre\ and   results in
\teff=17500~K, somewhat higher than the other values. This is a well known
phenomenon discussed e.g. by Hambly et al. (\cite{haro97}),  who derive 21000~K
for HR~6588 from the \ion{Si}{II}/\ion{Si}{III} ionization equilibrium as
opposed to 17700~K from Str\"omgren photometry. 

For the further analysis we use the  values derived from Str\"omgren photometry
and Balmer lines, as the  temperature information is more reliably described by
Str\"omgren  photometry at the low gravities involved, and keep the helium 
abundance fixed at \loghe\ = $-$1.00.

\begin{figure} 
\includegraphics{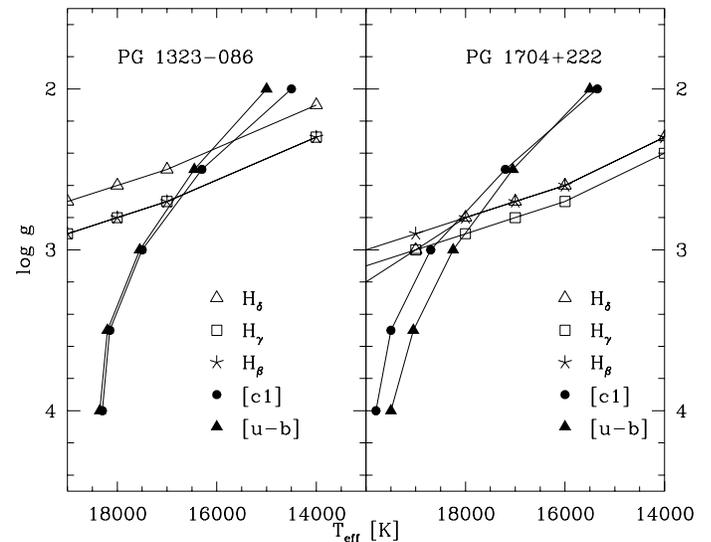}
\vspace{6.5cm}
\caption[]{The range of physical parameters resulting from Str\"omgren 
photometry (Wesemael et al., \cite{wese92}) and from fits to the Balmer lines. }
\end{figure}

\begin{table}
\begin{tabular}{|l|lll|}
\hline
Name &  T$_{\rm eff}$ & log g & \loghe\ \\
     &  [K] & & \\
\hline
\pgdre          & 16000$^1$ & 2.50$^1$ & $-$1.00$^4$ \\
               & 15700$^2$ & 2.35$^2$ & $-$0.83$^2$  \\ 
               & 17500$^3$ & 2.65$^3$ & $-$1.00$^4$  \\ 
\pgsie         & 18000$^1$ & 2.80$^1$ & $-$1.00$^4$ \\
               & 17200$^2$ & 2.65$^2$ & $-$0.94$^2$  \\ 
HR 6588$^5$    & 17700$^1$ & 3.90$^1$ & $-$1.23  \\
\hline
\end{tabular}\\
$^1$ from Str\"omgren photometry and Balmer lines\\
$^2$ from Balmer and \ion{He}{I} lines only\\
$^3$ \teff\ from \ion{Si}{II}/\ion{Si}{III}, \logg\ from the Balmer lines\\
$^4$ He abundance kept fixed\\
$^5$ Hambly et al. (\cite{haro97})\\
\caption{The physical parameters for the programme stars and the calibration 
star.}
\end{table}

In Fig. 3 the physical parameters are compared to post-AGB tracks from
Sch\"onberner (\cite{scho83}) and Bl\"ocker \& Sch\"onberner (\cite{blsc90}).
Also shown are  post-AGB stars analysed elsewhere (McCausland et al.,
\cite{mcco92}, Conlon et al., \cite{codu93}, and references therein). It can be
clearly  seen that \pgdre\ and \pgsie\ are well described by post-AGB tracks 
with masses between 0.546~\Msolar\ and 0.565~\Msolar .

\begin{figure}  \includegraphics{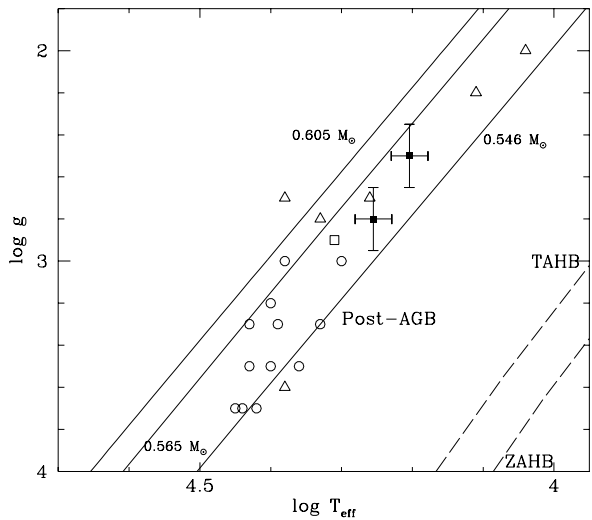} \vspace{6.5cm} \caption[]{The physical parameters of the
programme stars compared to  post-AGB tracks (solid lines)  by Sch\"onberner
(\cite{scho83}; 0.546~\Msolar, 0.565~\Msolar) and  Bl\"ocker \& Sch\"onberner
(\cite{blsc90}, 0.605~\Msolar). The dashed lines  show horizontal branch models
by Dorman et al. (\cite{dorm93}). Also marked  are the positions of possible
post-AGB stars analysed by other groups: Conlon  et al. (\cite{codu93},
circles, no abundance analyses);  Napiwotzki et al. (\cite{nahe94}, square);
McCausland et al. (\cite{mcco92}, triangles).} 
\end{figure}

\begin{table*}
\begin{tabular}{|l|llrr|r||l|llrr|r|}
\hline
Name & Ion & $\lambda$ & $\log gf$& W$_\lambda$ & $\log \epsilon$ &
Name & Ion & $\lambda$ & $\log gf$& W$_\lambda$ & $\log \epsilon$\\
     &         &  [\AA ]   & &  [m\AA ]    & &
     &         &  [\AA ]   & &  [m\AA ]    & \\
\hline
 \pgdre & \ion{C}{II} & 4267.02 &  0.559 &    &      & 
  HR 6588 & \ion{C}{II} & 4267.02 & 0.559 & &  \\
 &               & 4267.27 &  0.734 & $<$17 & $<$5.99  & 
        &      & 4267.27 &  0.734 & 195 & 8.09  \\
 & \ion{N}{II}   & 3995.00 &  0.225 & 20 & 6.77 & 
        &      & 4411.20 &  0.517 & &  \\
 & \ion{O}{II}   & 4069.62 &  0.157 &    &      & 
        &      & 4411.52 &  0.672 & 21 & 8.27 \\
 &               & 4069.89 &  0.365 & 15 & 7.89 & 
        &      & 6578.03 & -0.040 & 69 & 7.92 \\
 &               & 4072.16 &  0.546 & 13 & 7.86 & 
        &      & 6582.85 & -0.340 & 67 & 8.08 \\
 &               & 4349.43 &  0.085 &  6 & 7.32 & 
        & \ion{N}{II} & 3995.00 &  0.225 & 39 & 7.71 \\
 & \ion{Mg}{II}  & 4481.13 &  0.568 &    &      & 
        &      & 4447.03 &  0.238 & 18 & 7.82 \\
 &               & 4481.33 &  0.732 & 51 & 5.79 & 
        &      & 4601.48 & -0.385 & 21 & 8.04 \\
 & \ion{Al}{III} & 4528.91 & $-$0.294 &  &      & 
        &      & 4607.16 & -0.483 & 13 & 7.80 \\
 &               & 4529.20 &  0.660 & 20 & 5.57 & 
        &      & 4613.87 & -0.607 & 10 & 7.79 \\
 & \ion{Si}{II}  & 4128.07 &  0.369 & 27 & 5.97 & 
        & \ion{O}{II} & 4069.89 & 0.157 & & \\
 &               & 4130.89 &  0.545 & 31 & 5.87 & 
        &      & 4069.89 &  0.365 & 29 & 8.69 \\
 & \ion{Si}{III} & 4552.62 &  0.283 & 35 & 6.36 & 
        &      & 4072.16 &  0.546 & 21 & 8.68 \\
 &               & 4567.82 &  0.061 & 24 & 6.33 & 
        &      & 4414.90 &  0.211 & 28 & 8.75 \\
 & & & & & &  &      & 4416.97 & -0.041 & 24 & 8.84 \\
\pgsie & \ion{C}{II} & 4267.02 &  0.559   &    &      & 
        &      & 4590.97 &  0.346 & 11 & 8.64 \\
 &              & 4267.27 &  0.734   & 60 & 6.52 & 
        &      & 4595.96 & -1.037 & & \\
 &              & 6578.03 & $-$0.040 & 60 & 6.77 & 
        &      & 4596.18 &  0.196 & 14 & 8.91 \\
 &              & 6582.85 & $-$0.340 & 59 & 7.04 & 
        &      & 4649.14 &  0.343 & 37 & 8.76 \\
 & \ion{N}{II}  & 3995.00 &  0.225   & 33 & 6.86 & 
        &      & 4650.84 & -0.331 & 17 & 8.81 \\
 & \ion{O}{II}  & 4069.62 &  0.157   &    &      &  
        & \ion{Mg}{II} & 4481.13 & 0.568 & & \\
 &              & 4069.89 &  0.365   & 15 & 7.89 & 
        &  & 4481.33 & 0.732 & 213 & 7.15 \\
 &              & 4349.43 &  0.085   &  9 & 7.59 & 
        & \ion{Al}{III} & 4512.54 & 0.405 & 20 & 6.31 \\
 &              & 4414.90 &  0.211   & 14 & 7.77 & 
        &      & 4528.91 & -0.294 & & \\
 & \ion{Mg}{II} & 4481.13 &  0.568   &    &      & 
        &      & 4529.20 &  0.660 & 38 & 6.38 \\
 &              & 4481.33 &  0.732   & 68 & 6.13 & 
        & \ion{Si}{II} & 4128.07 & 0.369 & 63 & 6.61 \\
 & \ion{Al}{III} & 4528.91 & $-$0.294 &  &      & 
        &      & 4130.89 &  0.545 & 74 & 6.57 \\
 &               & 4529.20 &  0.660 & $<$20 & $<$5.38 & 
        & \ion{Si}{III} & 4552.62 & 0.283 & 74 & 7.68 \\
 & \ion{Si}{II} & 4128.07 &  0.369   & 26 & 6.17 & 
        &      & 4567.82 &  0.061 & 56 & 7.63 \\
 &              & 4130.89 &  0.545   & 38 & 6.17 & 
        &      & 4574.76 & -0.416 & 33 & 7.60 \\
\hline
\end{tabular}
\caption{The measured equivalent widths and derived abundances for the 
programme stars and the reference star  HR~6588 ($\epsilon$ denotes the 
particle numbers of the respective element with log $\epsilon$ = 
log(X/H)+12.00). For HR~6588 we used the same microturbulent velocity as
Hambly et al. (1997, 5~km/s).}
\end{table*}

\section{Abundances}

By comparing the spectra of the two objects to spectra of the well known 
main-sequence B star HR~6588 
(that were obtained with the same equipment) we identified
lines of \ion{O}{II}, \ion{N}{II}, \ion{C}{II} (only \pgsie ),  \ion{Si}{II},
\ion{Si}{III} (only \pgdre ), \ion{Mg}{II}, and \ion{Al}{III}  (only \pgdre ).
We measured equivalent widths (resp. upper limits) for those lines in the
object spectra and in the spectrum of HR~6588. 

A differential abundance analysis is performed using HR~6588 as a reference 
star. This obviously is not an ideal choice, since HR~6588 is a main  sequence
star. Although its \teff\ is similar to those of our programme  stars, its
gravity is considerably higher. 

From model atmospheres for the appropriate values of effective temperature and 
surface gravity (see Table~1) we calculated curves of growth for the elements 
mentioned above, from which we then  derived abundances. The number of spectral
lines of any ion is insufficient  to derive microturbulent velocities for the
programme stars.  Therefore  we adopted a  value of 15~km/sec, which is in good
agreement with microturbulent  velocities derived for other post-AGB stars
(McCausland et al., \cite{mcco92}, see also Table~4). A change of 5~km/sec in
the microturbulent velocity results typically in a  change of 0.05 dex or less
in abundance.  Even if we decrease the microturbulent  velocity to 5
km/sec the abundances increase by 0.08 to 0.2 dex only. A change in \teff\ by
1000~K results in abundance changes of less than 0.1~dex (except for
\ion{O}{II}: 0.2~dex). Table~2 lists the abundances derived from individual
lines for each star. 

Using the abundances derived from the same lines in the spectrum of
HR~6588  we determined abundances relative to HR~6588 on a line by line basis.
Most notably the abundances derived from \ion{Si}{II} and  \ion{Si}{III} lines
in \pgdre\ and HR~6588 disagree by 0.43~dex and 1.05~dex, respectively,
underlining the need for a differential abundance analysis. 

Among the many abundance analyses of HR~6588 in the literature we choose  the
recent work by Hambly et al. (\cite{haro97}) to convert the relative abundances
(averaged if possible) into absolute ones, since the authors used the same
calibration of  Str\"omgren colours to derive \teff\ as we did.  The results
are listed in Table~3 and plotted in Fig.~4.  The silicon abundance of \pgdre\
as derived from \ion{Si}{II} now agrees with that derived from the
\ion{Si}{III} lines. We adopt $\pm$0.3~dex as a typical error for all
abundances. 

All measured metals are heavily depleted with respect to solar  abundances,
carbon showing the largest depletion in both stars, nitrogen  and oxygen the
lowest depletions. The abundance patterns resemble very  closely that of
\bddre\ (Napiwotzki et al., \cite{nahe94}), a central star of  a halo
planetary nebula (halo CSPN), as well as those of PHL~1580 and  PHL~147, two
post-AGB stars analysed by Conlon et al. (\cite{codu91})  (see Fig.~4). Note,
however, that other B type PAGB stars  show a much larger scatter in abundance
patterns (see Table~4).

\begin{table}
\begin{tabular}{|ll|rrr|}
\hline
 name & ion & \multicolumn{3}{c|}{mean abundance} \\
      &     &  absolute & solar & relative to solar \\
\hline
\pgdre      & \ion{C}{II}   & $<$6.16 & 8.58 & $<-$2.42 \\
            & \ion{N}{II}   & 6.93 & 8.05 & $-$1.12 \\
            & \ion{O}{II}   & 7.98 & 8.93 & $-$0.95 \\
            & \ion{Mg}{II}  & 5.87 & 7.58 & $-$1.71 \\
            & \ion{Al}{III} & 5.42 & 6.47 & $-$1.05 \\
            & \ion{Si}{II}  & 6.17 & 7.55 & $-$1.38 \\
            & \ion{Si}{III} & 6.20 & 7.55 & $-$1.35 \\
\hline
\pgsie      & \ion{C}{II}   & 7.01 & 8.58 & $-$1.57 \\
            & \ion{N}{II}   & 7.02 & 8.05 & $-$1.03 \\
            & \ion{O}{II}   & 7.90 & 8.93 & $-$1.03 \\
            & \ion{Mg}{II}  & 6.21 & 7.58 & $-$1.37 \\
            & \ion{Al}{III} & $<$5.23 & 6.47 & $<-$1.24 \\
            & \ion{Si}{II}  & 6.42 & 7.55 & $-$1.13 \\
\hline
\end{tabular}
\caption{The mean abundances of the programme stars }
\end{table}

\begin{figure}  \includegraphics{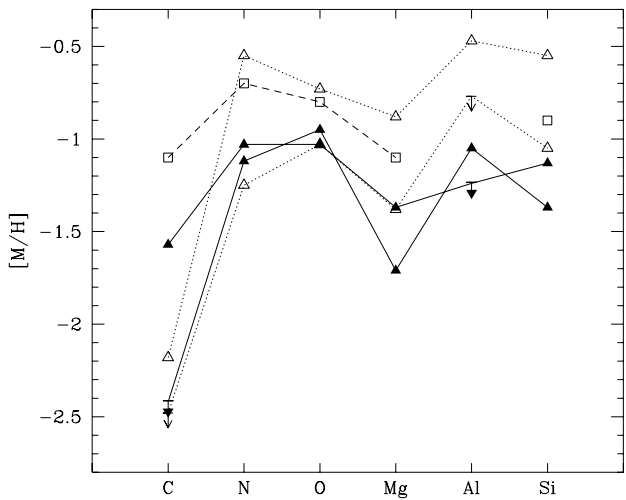} \vspace{6.5cm} \caption[]{The abundances determined in
this paper (filled symbols) compared to abundances of PAGB stars with similar
abundance patterns (open triangles: PHL~174, PHL~1580, Conlon et al.,
\cite{codu91};  open square: \bddre , Napiwotzki et al., \cite{nahe94}). The C
abundances relying only on the \ion{C}{II} 4267~\AA\ line might be
underestimated due to NLTE effects (Eber \& Butler, \cite{ebbu88}).}
\end{figure}

\begin{table*}
\begin{tabular}{|l|rr|rrrrrl|l|}
\hline
name     & \teff & \logg & [C/H] & [N/H] & [O/H] & [Mg/H] & [Al/H] & [Si/H]
 & $\xi_{micro}$\\
 & [K] & & & & & & & & [km/sec]\\
\hline
\pgdre    & 16000 & 2.5 & $<-$2.42 & $-$1.12 & $-$0.95 & $-$1.71 & 
$-$1.05  & $-$1.37 & 15$^1$\\
\pgsie    & 18000 & 2.8 &  $-$1.57 & $-$1.03 & $-$1.03 & $-$1.37 & 
$<-$1.24 & $-$1.13 & 15$^1$\\
\hline
LS~IV-4.01$^3$ & 11000 & 2.0 & $<-$1.38 &  --     &   --    & $-$1.98 &  
 --     & $-$2.05 & 18 \\ 
LB~3193$^3$  & 12900 & 2.2 & $<-$2.08 &  --     &   --    & $-$2.08 & 
  --     & $-$2.15 & 15$^1$\\ 
PHL~174$^2$ & 18200 & 2.7 & $<-$2.48 & $-$1.25 & $-$1.03 & $-$1.38 & 
$<-$0.77 & $-$1.05 & 10\\
\bddre $^4$ & 20200 & 2.9 & $-$1.1\hspace*{0.5em} & $-$0.7\hspace*{0.5em} 
& $-$0.8\hspace*{0.5em} & $-$1.1\hspace*{0.5em} & & $-$0.9\hspace*{0.5em} & 
15$^1$\\
LB~3219$^3$      & 21400 & 2.8 & $-$1.88  & $-$0.45 & $-$1.33 & $-$0.58 & 
$<-$1.27 & $-$0.95 & 16\\ 
LS~IV-12.111$^3$ & 24000 & 2.7 & $-$1.88  & $-$0.25 & $-$0.13 & $-$0.28 & 
$<-$1.07 & +0.05 & 15$^1$\\ 
PHL~1580$^2$ & 24000 & 3.6 & $-$2.18  & $-$0.55 & $-$0.73 & $-$0.88 &
  $-$0.47   & $-$0.55 & 8\\ 
\hline
\end{tabular}\\
\begin{tabular}{l}
$^1$ set to this value for the abundance analysis (not determined)\\
$^2$ Conlon et al. (\cite{codu91})\\
$^3$ McCausland et al. (\cite{mcco92})\\
$^4$ Napiwotzki et al. (\cite{nahe94})\\
\end{tabular}\\
\caption[]{The abundances of \pgdre\ and \pgsie\ compared to those of 
other PAGB stars of spectral type B.}
\end{table*}

\section{Kinematics} Using the derived physical parameters and the photometric
data by Wesemael et al. (\cite{wese92}) together with theoretical fluxes and
colours from Kurucz (\cite{kuru92})  and an assumed mass of 0.55~\Msolar we can
estimate the distances of the  two stars. The results are listed in Table~5.
The error of 23\% in distance is  dominated by the error in \logg\ (Moehler et
al., \cite{mohe90b}), which we  estimate to be 0.15 dex. Thejll et al.
(\cite{thfl97})  also measured proper motions for both stars, which we use to
derive  galactocentric velocities according to Johnson \& Soderblom
(\cite{joso87}). We used 8~kpc for the galactocentric distance of the Sun, (10,
15, 8) [km/s] for (U$_\odot$, V$_\odot$, W$_\odot$) and 225~km/s for  the
galactic rotation velocity at the place of the Sun. We also derived the
velocities of \pgsie\ along the axis pointing from  the galactic center to its
current position (within the galactic plane,  ${\rm \Pi}$) and  along the
galactic rotation at its place (${\rm \Theta}$). From the value of  ${\rm
\Theta}$ we note that \pgsie\ does not participate in the galactic  rotation.
The error of the proper motion of \pgdre\ is so large that  the resulting space
velocity is inconclusive and therefore not listed in  Table~5.

\begin{table*} \begin{tabular}{|l|rr|rr|rrr|rrr|rr|} \hline Name & l & b & d &
z & v$_{\rm r,hel.}$ & ${\rm \mu_\alpha}$ &  ${\rm \mu_\delta}$ & U$_{\rm GSR}$
& V$_{\rm GSR}$ & W$_{\rm GSR}$ &  ${\rm \Pi}$ & ${\rm \Theta}$ \\ & [$^\circ$]
& [$^\circ$] & [kpc] & [kpc] & [km/s] & [mas/yr] & [mas/yr] &  [km/s] & [km/s]
& [km/s] & [km/s] & [km/s] \\ \hline \pgdre  & 317.1 & +53.1 & 15.8 & 12.6 &
$-$46 & $-$3.9  & +0.8 & -- & -- & -- & -- & -- \\ &  &  & $\pm$3.7  & $\pm$3.0
& $\pm$15 & $\pm$ 4.0 & $\pm$ 4.9 & -- & -- & -- & -- & -- \\ & & & & & & & & &
& & & \\ \pgsie  & 42.9 & +32.4 & 6.9 & 3.6 & $-$38 & $-$8.4 & 0 & $-$53  & +57
& +209 & +78 & +1 \\ & & & $\pm$1.6 & $\pm$0.9 & $\pm$15 & $\pm$3.1 & -- &
$\pm$19 & $\pm$72 &  $\pm$98 & $\pm$54 & $\pm$51 \\ \hline \end{tabular}
\caption[]{Kinematic data for the two programme stars. U points from the  sun
to the galactic center, V towards l = 90$^\circ$, b = 0$^\circ$, W  towards the
galactic north pole. ${\rm \Pi}$ and ${\rm \Theta}$ are defined  to point from
the galactic center to the star (within the galactic plane)  and along the
galactic rotation at the place of the star, respectively.} 
\end{table*}

\section{Conclusions}

\pgdre\ and \pgsie\ are giant stars with normal helium abundances. Both stars
share a characteristic abundance pattern with  \bddre , the central star of a
halo planetary nebula, and with  two high galactic latitude B-giants discussed
by Conlon et al. (\cite{codu91}), most notably the strong carbon deficiency. 
Depletions of N, O, Mg, Al, and Si by 1 dex or more in \pgdre\ and \pgsie\
point towards a low metallicity and, hence, membership of an old stellar
population,  while they are not compatible with the values found for young B
stars.  The positions of both stars in the (\teff, log\,g) diagram are
consistent with evolutionary tracks starting from the AGB for masses of about
0.55~\Msolar. We therefore conclude that both stars are in the post-AGB stage
of their  evolution.

The similarity of the abundance patterns to that of \bddre\ suggests that  the
successors of the B-type PAGB stars can be found amongst the halo CSPNe as
already suggested by McCausland et al. (\cite{mcco92}). Potential progenitors
could be found among the optically bright post-AGB stars discussed by van
Winckel (\cite{vawi97}). However, van Winckel finds only one star, HD\,107369,
for which the abundance pattern comes close to that   of B-type PAGB stars. 
The CNO abundances of this star are  consistent with being the
product of CN cycling. In the B-type PAGB stars, however, the C/N ratio can 
be explained by CN cycling, but not the N/O ratio.

Conlon et al. (\cite{codu91}), McCausland el. (\cite{mcco92}) and Napiwotzki et
al. (\cite{nahe94}) discussed various  scenarios to explain the abundance
pattern of B-type PAGB stars  in terms of nuclear processing and  subsequent
dredge-up, but no conclusion could be reached.  The high [O/Fe] ratio
(+1.2$\pm$0.5) observed in \bddre\ prompted  Napiwotzki et al. (\cite{nahe94})
to suggest that its abundance pattern could be affected by gas-dust separation 
invoked by Bond (\cite{bond91}) to explain the extreme [O/Fe]~$>$~4 ratio 
found in two peculiar PAGB stars of spectral type F (Lambert et al., 
\cite{lahi88} and Waelkens et al., \cite{wawi91}). Therefore, the determination
of the iron abundance in the B-type PAGB  stars is a prerequisite for an
interpretation of their abundance patterns.  

According to the kinematic behaviour of \pgsie\ and the large distance  from
the galactic plane for \pgdre\ both stars are probably Population~II  objects. 

\acknowledgements

We want to thank the staff from the Calar Alto observatory for their support
during our observations, Michael Lemke for his help  with the abundance
 analysis and the referee Detlef Sch\"onberner for his  valuable remarks and 
suggestions. SM acknowledges financial support from the Deutsche 
Forschungsgemeinschaft through grants Mo 602/3 and Mo 602/4 and from the DARA 
under grant 50~OR~96029-ZA. This research has made use of the SIMBAD data base, 
operated at CDS, Strasbourg, France.

\end{document}